\documentclass{article}      

\usepackage{epsfig}

\title{q-Analogue of  Shock Soliton Solution}

\author{Sengul Nalci and Oktay K. Pashaev \\Department of Mathematics, Izmir Institute of Technology \\ Urla-Izmir, 35430, Turkey}

\begin{document}

\maketitle


\begin{abstract}
By using Jackson's q-exponential function we introduce the generating function, the recursive formulas and the second order q-differential
equation for the q-Hermite polynomials.
 This allows us to solve the q-heat equation in terms of
q-Kampe de Feriet polynomials with arbitrary N moving zeroes, and to find operator solution for the Initial Value Problem for the q-heat equation. By the q-analog of
the Cole-Hopf transformation we construct the q-Burgers type nonlinear heat equation with quadratic dispersion and the cubic nonlinearity. In $q \rightarrow 1$ limit it
reduces to
the standard Burgers equation. Exact solutions for the q-Burgers equation in the form of moving poles,  singular and regular q-shock soliton solutions are found.

\end{abstract}

\section{Introduction}

It is well known that the Burgers' equation in one dimension could be linearized by the Cole-Hopf transformation in terms of the linear heat equation.
It allows one to solve the initial value problem for the Burgers equation and to get exact solutions in
the form of shock solitons and describe their scattering. In the present paper we study the q-differential Burgers type  equation  with quadratic dispersion and the cubic
nonlinearity, and find its linearization in terms of the q-heat equation. In terms of the Jackson's q-exponential function we introduce the q-Hermite and q-Kampe-de
Feriet polynomials, representing moving poles solution for the q-Burgers equation. Then we derive the operator solution of the initial value problem for the q-Burgers
equation in terms of the IVP for the q-heat equation. We find solutions of our q-Burgers type equation in the form of singular and regular q-shock solitons. It turns out
that static q-shock soliton solution shows remarkable self-similarity property in space coordinate $x$.

\section{q-Exponential Function}

The q-number corresponding to the ordinary number $n$ is defined as, \cite{Kac et al.},
\begin{equation}[n]_q= \frac{q^n-1}{q-1},\end{equation}where $q$ is a parameter, so that $n$ is the limit of $[n]_q$ as $q\rightarrow 1$. A few examples of
q-numbers are given here:
$[0]_q=0$, $[1]_q=1$, $[2]_q=1+q$, $[3]_q=1+q+q^2$.
In terms of these q-numbers, the Jackson q-exponential function is defined as \begin{equation}e_q(x)=\sum^{\infty}_{n=0}\frac{x^n}{[n]_q!}.\end{equation}For $q>1$ it is
entire function of $x$ and
when $q\rightarrow 1$ it reduces to the standard exponential function $e^x$.
The q-exponential function can also be expressed in terms of infinite product
\begin{equation}e_q(x)=
\prod^\infty_{n=0}\frac{1}{\left(1-(1-q)q^n x\right)} =
\frac{1}{\left(1-(1-q)x\right)^\infty_q},\end{equation} when $q<1$ and
\begin{equation}e_q(x)=
\prod^\infty_{n=0}{\left(1+(1-\frac{1}{q})\frac{1}{q^n} x\right)}=
{\left(1+(1-\frac{1}{q})x\right)^\infty_{1/q}},\end{equation} when $q>1$.
Thus, the q-exponential function for $q<1$ has infinite set of poles at \begin{equation}x_n=\frac{1}{q^n(1-q)},\,\,\,\,\,\,\, n=0,1,..\label{polesq}\end{equation}
and for $q>1$ the infinite set of zeros at \begin{equation}x_n=-\frac{q^{n+1}}{(q-1)},\,\,\,\,\,\,\, n=0,1,..\label{zerosq}\end{equation}
The q-derivative is defined as
\begin{equation}D^x_qf(x)=\frac{f(qx)-f(x)}{(q-1)x},\end{equation}
and when $q\rightarrow1$ it reduces to the standard derivative  $D^x_qf(x)\rightarrow f'(x)$.
Using the definition of the $q$-derivative one can easily see that
\begin{eqnarray} &&D^x_q(ax^n) = a[n]_qx^{n-1},\\
&&D^x_qe_q(ax)=ae_q(ax).\end{eqnarray}

\section{q-Hermite Polynomials}

We define the q-Hermite polynomials according to the generating function
\begin{equation} e_q(-t^2) e_q([2]_q t x) = \sum^{\infty}_{n=0} H_n(x;q) \frac{t^n}{[n]_q!}.\label{genfunc}
 \end{equation}
From this generating function we have the special values
\begin{eqnarray} &&H_{2n}(0;q) = (-1)^n \frac{[2n]_q!}{[n]_q!},\\
&&H_{2n+1}(0;q) = 0, \end{eqnarray}where $[n]_q!=[1]_q[2]_q...[n]_q$,
and the parity relation
\begin{equation} H_n (-x;q) = (-1)^n H_n (x;q). \end{equation}
By q-differentiating the generating function (\ref{genfunc})  according to $x$ and $t$ we have the recurrence relations
correspondingly

\begin{equation} D_x H_n (x;q) = [2]_q [n]_q H_{n-1}(x;q),\label{rec1}
\end{equation}
\begin{equation}
H_{n+1} (x;q) = [2]_q \,x  H_{n}(x;q) - [n]_q \,H_{n-1}(q x;q)\nonumber \\
- [n]_q \,q^{\frac{n+1}{2}} H_{n-1}(\sqrt{q} x;q).\label{rec2}
\end{equation}

Using operator
\begin{equation}M_q = q^{x \frac{d}{dx}},\end{equation}
so that
\begin{equation} M_q f(x) = f(q x),\end{equation}
relation (\ref{rec2}) can be rewritten as
\begin{equation}
H_{n+1} (x;q) = [2]_q \,x  H_{n}(x;q) \nonumber\\
- [n]_q (M_q + q^{\frac{n+1}{2}} M_{\sqrt{q}}) H_{n-1}(x;q).\label{rec3}
\end{equation}
Substituting (\ref{rec1}) to (\ref{rec3}) we get
\begin{equation}
H_{n+1} (x;q) = \left([2]_q \,x  - \frac{M_q + q^{\frac{n+1}{2}} M_{\sqrt{q}}}{[2]_q} D_x\right) H_{n}(x;q)\label{rec4}
\end{equation}
By the recursion, starting from $n=0$ and $H_0(x) = 1$  we have next representation for the q-Hermite polynomials
\begin{equation}
H_{n} (x;q) = \prod^n_{k=1}\left([2]_q \,x  - \frac{M_q + q^{\frac{k}{2}} M_{\sqrt{q}}}{[2]_q}  D_x\right)\cdot 1\label{rec5}
\end{equation}
We notice that the generating function and   the form of our q-Hermite polynomials are different from the known ones in the literature, \cite{Exton}, \cite{Cigler et
al.},
\cite{Rajkovic et al.}, \cite{Negro}.\\
In the above expression the operator
\begin{equation}
M_q + q^{\frac{n}{2}} M_{\sqrt{q}} =  2 q^{\frac{n}{4}}q^{\frac{3}{4}x \frac{d}{dx}} \cosh [(\ln q^{\frac{1}{4}})(x \frac{d}{dx} - n)]
\end{equation}
is expressible in terms of the q-spherical means as
\begin{equation}
\cosh [(\ln q ) x \frac{d}{dx}] f(x) = \frac{1}{2}(f(q x) + f(\frac{1}{q}x)).
\end{equation}

Using definition, \cite{Kac et al.},
\begin{eqnarray}
(x-a)^n_q = (x -  a)(x - q a)\cdot\cdot\cdot  (x - q^{n-1} a)
,\,\,n=1,2,.. \nonumber
\end{eqnarray}
which now we apply for operators, we should distinguish the direction of multiplication. We consider two cases
\begin{equation}
(x-a)^n_{q \,<} = (x -  a)(x - q a)\cdot\cdot\cdot (x - q^{n-1} a),\label{r}
\end{equation}
and
\begin{equation}
(x-a)^n_{q \,>} = (x - q^{n-1} a)\cdot\cdot\cdot (x - q a) (x -  a).\label{l}
\end{equation}
Then, we can rewrite (\ref{rec5}) shortly as
\begin{eqnarray}H_{n} (x;q) = \left( ([2]_q \,x - \frac{ M_q \,D_x}{[2]_q} )- q^{\frac{1}{2}} \frac{M_{\sqrt{q}}\, D_x}{[2]_q} \right)^n_{\sqrt{q}\, >} \cdot1
\label{rec6}\nonumber \end{eqnarray}
First few polynomials are
\begin{eqnarray} H_0 (x;q) = 1\nonumber ,\end{eqnarray}\begin{eqnarray}H_1 (x;q) = [2]_q \,x \nonumber,\end{eqnarray}
\begin{eqnarray} H_2(x;q) = [2]^2_q \, x^2 - [2]_q\nonumber,\end{eqnarray}
\begin{eqnarray} H_3(x;q) = [2]^3_q \,x^3 - [2]^2_q [3]_q \,x\nonumber, \end{eqnarray}
\begin{eqnarray} H_4(x;q) = [2]^4_q \,x^4 - [2]^2_q [3]_q [4]_q\,x^2 + [2]_q [3]_q [2]_{q^2} \nonumber .\end{eqnarray}
When $q\rightarrow 1$ these polynomials reduce to the standard Hermite polynomials.

\subsection{q-Differential Equation}

Applying $D_x$ to both sides of  (\ref{rec4}) and using recurrence formula (\ref{rec1}) we
get q-differential equation for q-Hermite polynomials
\begin{eqnarray} \frac{1}{[2]_q} D_x (M_q + q^{\frac{n+1}{2}} M_{\sqrt{q}}) D_x H_n (x;q) \nonumber -
[2]_q q x D_x H_n(x;q) + [2]_q [n]_q q H_n(x;q) = 0
\nonumber .\end{eqnarray}

\subsection{Operator Representation}

\newtheorem{prop}{Proposition}
\newtheorem{pf}{Proof}
\newtheorem{cor}{Corrollary}

\begin{prop} We have next identity

\begin{equation} e_q \left(-\frac{1}{[2]_q} D^2_x \right) e_q ([2]_q x t) = e_q (-t^2) e_q ([2]_q x t)\label{prop1}.\end{equation}
\end{prop}

\begin{pf}
By q- differentiating the q-exponential function  in $x$
\begin{equation} D^n_x e_q ([2]_q x t) = ([2]t)^n e_q ([2]_q x t),\end{equation}
and combining then to the sum
\begin{equation} \sum_{n=0}^{\infty} \frac{a^n}{[n]_q!} D^{2n}_x e_q ([2]_q x t) =  \sum_{n=0}^{\infty} \frac{[2]_q^n a^n t^{2n}}{[n]_q!}\,\, e_q ([2]_q x
t),\end{equation}
we have relation
\begin{equation} e_q(a D^2_x) e_q ([2]_q x t) = e_q ([2]_q a t^2) e_q ([2]_q x t).\end{equation}
By choosing $a = - 1/[2]_q$ we get
\begin{equation} e_q \left(-\frac{1}{[2]_q} D^2_x \right) e_q ([2]_q x t) = e_q (-t^2) e_q ([2]_q x t).\end{equation}
\end{pf}
\begin{prop}The next identity is valid

\begin{equation}H_n (x; q) =  [2]^n e_q \left(-\frac{1}{[2]_q} D^2_x \right) x^n \label{prop2}.\end{equation}
\end{prop}

\begin{pf} The right hand side of (\ref{prop1}) is the generating function for q-Hermite polynomials. Hence expanding both sides
in $t$ we get the result.
\end{pf}

\begin{prop}
\begin{equation}
  e_q \left(-\frac{D^2_x}{[2]_q} \right) x^{n+1}= \nonumber \\ \frac{1}{[2]_q}\left([2]_q \,x - \frac{(M_q + q^{\frac{n+1}{2}} M_{\sqrt{q}}) D_x}{[2]_q} \right) e_q
\left(-\frac{D^2_x}{[2]_q} \right) x^{n}.\nonumber
\label{prop3}\end{equation}
\end{prop}
\begin{pf} we use (\ref{prop2}) and relation (\ref{rec4})
.\end{pf}

\begin{cor} If function $f(x)$ is analytic and expandable to power series $f(x) = \sum^\infty_{n=0} a_n x^n$ then
we have next q-Hermite series
\begin{equation} e_q \left(-\frac{1}{[2]_q} D^2_x \right) f(x) = \sum^\infty_{n=0} a_n \frac{H_n(x;q)}{[2]_q^n}.\end{equation}
\end{cor}

\section{q- Kampe-de Feriet Polynomials}

We define q-Kampe-de Feriet polynomials as
\begin{equation}
H_n(x, t;q)= (-\nu t)^{\frac{n}{2}} H_n\left( \frac{x}{[2]_q \sqrt{-\nu t}}\right)
,\end{equation}
and from (\ref{rec4}) we have the next recursion formula 
\begin{eqnarray}
H_{n+1} (x, t;q) = \left(x  + (M_q + q^{\frac{n+1}{2}} M_{\sqrt{q}}) \nu t D_x\right) H_{n}(x, t;q)\label{recKF}\nonumber
.\end{eqnarray}
By recursion it gives
\begin{equation}
H_{n} (x,t;q) = \prod^n_{k=1}\left( x  +  (M_q + q^{\frac{k}{2}} M_{\sqrt{q}})\nu t D_x\right)\cdot 1\label{rec1 KF}
\end{equation}
or by notation (\ref{l})
\begin{eqnarray}
H_{n} (x,t;q) = \left( (x  + M_q \,\nu t \,D_x ) + q^{\frac{1}{2}}  M_{\sqrt{q}}\,\nu t D_x\right)^n_{\sqrt{q}\, >}\cdot 1 \label{rec2 KF}\nonumber
\end{eqnarray}
First few polynomials are
\begin{eqnarray} H_0 (x, t;q) = 1\nonumber,\end{eqnarray} \begin{eqnarray}H_1 (x,t ;q) = x\nonumber,\end{eqnarray}
\begin{eqnarray} H_2(x,t ;q) =  x^2 + [2]_q \, \nu t\nonumber,\end{eqnarray}
\begin{eqnarray} H_3(x,t ;q) = x^3 + [2]_q [3]_q \,\nu t \,x\nonumber ,\end{eqnarray}
\begin{eqnarray} H_4(x,t ;q) = x^4 + [3]_q [4]_q\,\nu t \,x^2 + [2]_q [3]_q [2]_{q^2} \nu^2 t^2\nonumber.\end{eqnarray}

\section{q-Heat Equation}

We consider the q-heat equation
\begin{equation} (D_t - \nu D^2_x) \phi(x,t) = 0 \label{qheat}.\end{equation}
Solution of this equation expanded in terms of parameter $k$
\begin{equation} \phi(x,t) = e_q(\nu k^2 t)e_q(k x) = \sum^\infty_{n=0} \frac{k^n}{[n]!} H_n(x,t;q),\end{equation}
gives the set of q-Kampe-de Feriet polynomial solutions for the equation.
Then we find time evolution of zeroes for these solutions in terms of zeroes $z_k(n,q)$ of q-Hermite polynomials,
\begin{equation}H_n(z_k(n,q),q) = 0 ,\end{equation}
so that

\begin{equation} x_k(t) = [2] z_k(n,q) \sqrt{-\nu t}.\label{zerosmotion}\end{equation}
For n=2 we have two zeros determined by q-numbers,
\begin{equation}x_1(t) = {\sqrt{[2]_q}} \sqrt{-\nu t}\\, \end{equation}
\begin{equation}x_2(t) = -{\sqrt{[2]_q}} \sqrt{-\nu t}\\, \end{equation}
and moving in opposite directions according to (\ref{zerosmotion}). For n=3 we have zeros determined by q-numbers,
\begin{eqnarray}x_1(t) &=& - \sqrt{{[3]_q!}}\sqrt{-\nu t} , \\
x_2(t) &=& 0 , \\
x_3(t) &=& \sqrt{{[3]_q!}}\sqrt{-\nu t} , \end{eqnarray}
two of which are moving in opposite direction according to (\ref{zerosmotion}) and one is in the rest.

\section{Evolution Operator}

Following similar calculations as in Proposition I we have next relation
\begin{equation} e_q \left(\nu t D^2_x \right) e_q (k x) = e_q ( \nu t k^2) e_q ( k x)\label{propKF}.\end{equation}
The right hand side of this expression is the plane wave type solution of the q-heat equation
\begin{equation} (D_t - \nu D^2_x) \phi(x,t) = 0 \label{qheat}.\end{equation}
Expanding both sides in power series in $k$ we get
 q-Kampe de Ferie polynomial solutions of this equation
\begin{equation} H_n(x,t;q) = e_q \left(\nu t D^2_x \right) x^n.\end{equation}

Consider an arbitrary analytic function $f(x) = \sum^\infty_{n=0} a_n x^n$, then function
\begin{eqnarray} f(x,t) = e_q \left(\nu t D^2_x \right) f(x) &=& \sum^\infty_{n=0} a_n  e_q \left(\nu t D^2_x \right) x^n \\ &=& \sum^\infty_{n=0} a_n
H_n(x,t;q),\end{eqnarray}
is a time dependent solution of the q-heat equation (\ref{qheat}).

According to this we have the evolution operator for the q-heat equation as
\begin{equation} U(t) = e_q \left(\nu t D^2_x \right).\end{equation}
It allows us to solve the initial value problem
\begin{eqnarray}(D_t - \nu D^2_x) \phi(x,t) &=& 0 ,\\
 \phi(x,0^+) &=& f(x),\end{eqnarray}
in the form
\begin{equation} \phi(x,t) =  e_q \left(\nu t D^2_x \right) \phi(x,0^+) = e_q \left(\nu t D^2_x \right) f(x).\label{IVP}\end{equation}

\section{q-Burgers' Type Equation}

Let us consider the q-Cole-Hopf transformation
\begin{equation} u(x,t) = -2 \nu \frac{D_x\phi(x,t)}{\phi(x,t)} ,\label{C H}\end{equation}
then $u(x,t)$ satisfies the q-Burgers' type Equation with quadratic dispersion and cubic nonlinearity
$$\begin{array}{l}D_tu(x,t)- D^2_xu(x,t)=\frac{1}{2} \left[(u(x,qt)- u(x,t)M^x_q)D_xu(x,t)\right]-\\ \\\frac{1}{2}\left[D_x\left(u(qx,t)u(x,t)\right)\right]+
\frac{1}{4} \left[u(q^2x,t)-u(x,qt)\right]u(qx,t)u(x,t).\end{array}$$
When $q \rightarrow 1$ it reduces to the standars Burgers' Equation.
\subsection{I.V.P. for q-Burgers' Type Equation }
Substituting the operator solution (\ref{IVP}) to (\ref{C H}) we find operator solution for the q-Burgers type equation in the form
\begin{equation} u(x,t) = -2 \nu \frac{e_q \left(\nu t D^2_x \right) D_x f(x)}{e_q \left(\nu t D^2_x \right)f(x)} .\label{operator Burgers}\end{equation}This solution
corresponds to the initial function
\begin{equation} u(x,0^+) = -2 \nu \frac{ D_x f(x)}{
f(x)} .\label{initial Burgers}\end{equation}
Thus, for arbitrary initial value problem for the q-Burgers equation with $u(x,0^+)=F(x)$ we need to solve the initial value problem for the q-heat equation with initial
function $f(x)$ satisfying the first order q-differential equation
\begin{equation} (D_x+\frac{1}{2\nu}F(x))f(x)=0.\end{equation}
\section{q-Shock soliton solution}

As a solution of q-heat equation we choose first  \begin{equation}\phi(x,t) = e_q \left(k^2t\right) e_q\left(kx\right),\end{equation}then we find
solution of the q-Burgers equation as a constant \begin{equation}u(x,t)=-2\nu k. \end{equation}
If we choose \begin{equation}\phi(x,t) =  e_q \left(k_1^2t\right) e_q\left(k_1x\right)+ e_q \left(k_2^2t\right) e_q\left(k_2x\right),\end{equation}
then we have the q-Shock soliton solution
\begin{equation}u(x,t) = -2\nu \frac{k_1 e_q \left(k_1^2t\right) e_q\left(k_1x\right)+k_2 e_q \left(k_2^2t\right) e_q\left(k_2x\right)}{e_q \left(k_1^2t\right)
e_q\left(k_1x\right)+ e_q \left(k_2^2t\right) e_q\left(k_2x\right)}.\label{qshock}\end{equation}
Due to zeroes of the q-exponential function this expression admits singularities for some values of parameters $k_1$ and $k_2$.
In Fig.1 we plot the singular q-shock soliton for $k_1 = 1$ and $k_2 = 10$ at time $t=0$.

\begin{figure}[h]
\begin{center}
\epsfig{figure=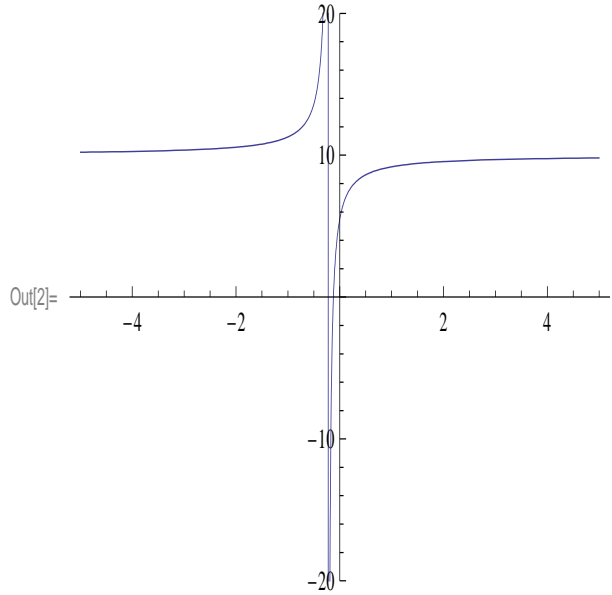,height=8cm,width=8cm}
\end{center}
\caption{Singular q-Shock Soliton} \label{qsss}
\end{figure}

However for some specific values  of the parameters we found the regular q-shock soliton solution.
We introduce the q-hyperbolic function \begin{equation}\cosh_q(x)=\frac{e_q(x)+e_q(-x)}{2} ,\end{equation}
or \begin{equation}\cosh_q(x)=\frac{1}{2}
\left( e_q(x)+\frac{1}{e_{\frac{1}{q}}(x)}\right),\end{equation}
then by using infinite product representation for q-exponential function we have

  \begin{eqnarray}\cosh_q(x)=\frac{1}{2}
\left( {\left(1+(1-\frac{1}{q})x\right)^\infty_{1/q}}+{\left(1-(1-\frac{1}{q})x\right)^\infty_{q}}
\right).\nonumber\end{eqnarray}
From (\ref{polesq}),(\ref{zerosq}) we find that zeroes of the first product are located on negative axis $x$, while for the second product on the positive axis $x$.
Therefore  the function has no zeros for real $x$ and $\cosh_q(0) =1$.

If $k_1 = 1$, and $k_2 = -1$, the time dependent factors in nominator and the denominator of (\ref{qshock}) cancel each other and  we have the stationary shock soliton
\begin{equation} u(x,t)=-2\nu \frac{e_q(x)-e_q(-x)}{e_q(x)+e_q(-x)}\equiv -2\nu\tanh_q(x).\end{equation}
Due to above consideration this function has no singularity on real axis and we have regular q-shock soliton.

In Fig.2, Fig.3 and Fig.4  we plot the regular q-shock soliton for $k_1 = 1$ and $k_2 = -1$  at different ranges of $x$.
It is remarkable fact that the structure of our shock soliton shows self-similarity property in space coordinate $x$.
Indeed at the ranges of parameter $x = 50,\,5000,\, 500 000$ the structure of shock looks almost the same.

\begin{figure}[h]
\begin{center}
\epsfig{figure=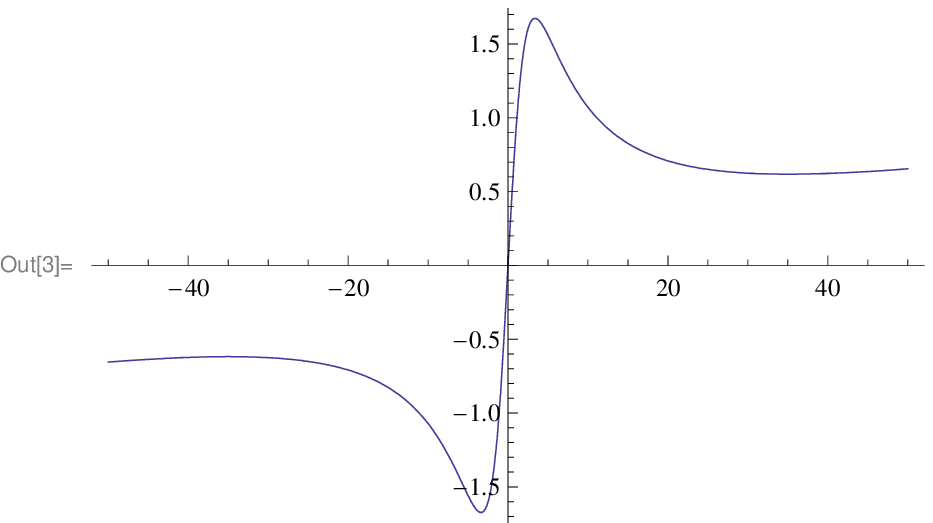,height=8cm,width=8cm}
\end{center}
\caption{Regular q-Shock Soliton For $k_1 = 1$, $k_2 = -1$, at range (-50, 50)} \label{q-shockregular}
\end{figure}

\begin{figure}[h]
\begin{center}
\epsfig{figure=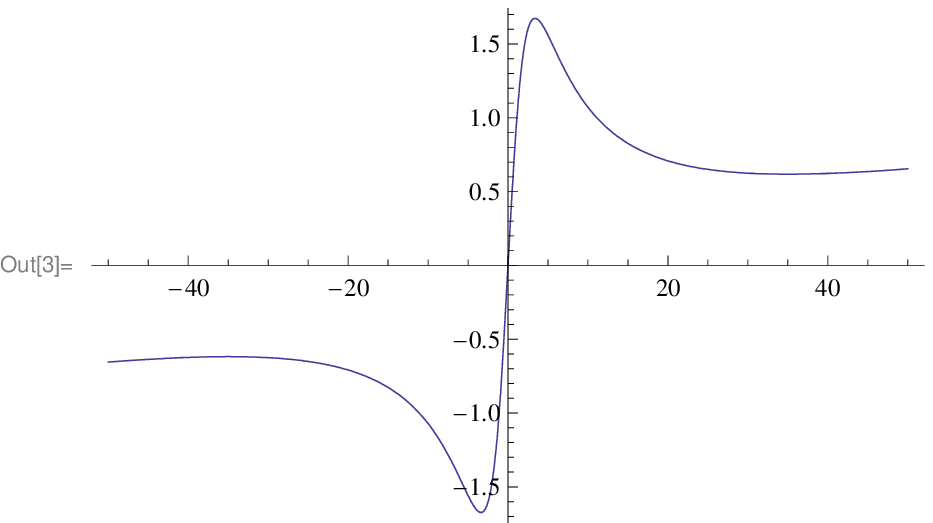,height=8cm,width=8cm}
\end{center}
\caption{The regular q-shock soliton for $k_1 = 1$, $k_2 = -1$ at range (-5000, 5000)} \label{q-shockregular5000}
\end{figure}

\begin{figure}[h]
\begin{center}
\epsfig{figure=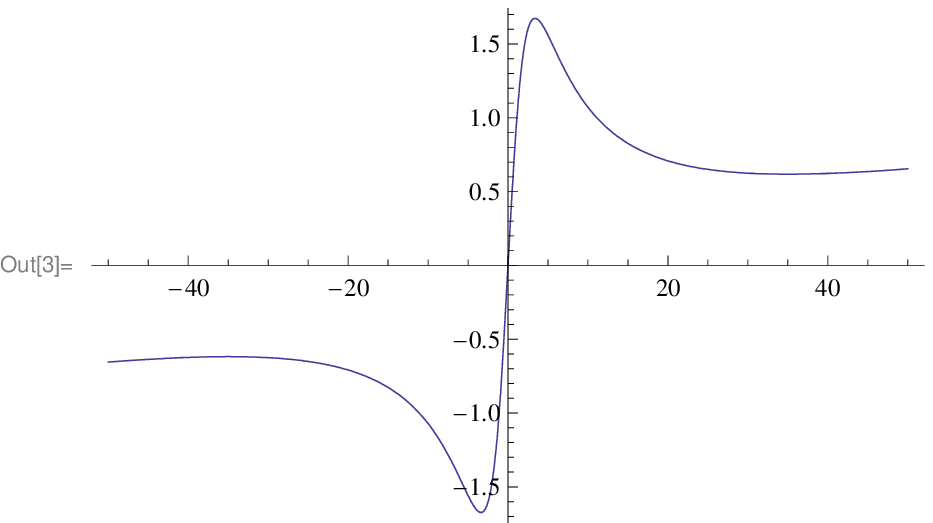,height=8cm,width=8cm}
\end{center}
\caption{The regular q-shock soliton for $k_1 = 1$, $k_2 = -1$  at range (-500000, 500000)} \label{q-shockregular50000}
\end{figure}

For the set of arbitrary numbers $k_1,...,k_N$
\begin{equation}\phi(x,t)= \sum^{N}_{n=1}e_q \left(k_n^2t\right) e_q\left(k_nx\right),\end{equation} we have multi-shock
solution in the form
\begin{equation}u(x,t) = -2 \nu\frac{\sum^{N}_{n=1}k_n e_q \left(k_n^2t\right) e_q\left(k_nx\right)}{\sum^{N}_{n=1}e_q \left(k_n^2t\right)
e_q\left(k_nx\right)}.\end{equation}

In general this solution admits several singularities.
To have regular multi-shock solution we can consider the even number of terms $N = 2k$ with opposite wave numbers.
When  $N = 4$ and $k_1 = 1$, $k_2 = -1$,$k_3 = 10$,$k_4 = -10$
we have q-multi-shock soliton solution,

\begin{eqnarray} u(x,t)=-2\nu\frac{e_q(t)\sinh_q(x)+ 10 e_q(100t)\sinh_q(10x)}{e_q(t)\cosh_q(x)+e_q(100t)\cosh_q(10x)}.\end{eqnarray}

In Fig. 5 we plot $N = 4$ case with values of the wave numbers $k_1 = 1$, $k_2 = -1$,$k_3 = 10$,$k_4 = -10$ at $t=0$.
To have regular solution for any time $t$ and given base $q$, we should choose proper numbers $k_i$ which are not in the form of power of $q$.
This question is under the study now.

\begin{figure}[h]
\begin{center}
\epsfig{figure=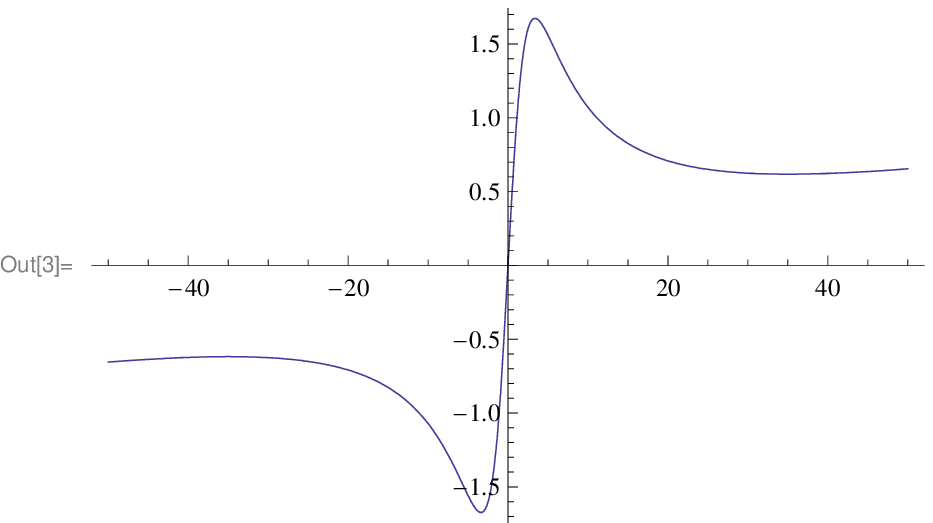,height=8cm,width=8cm}
\end{center}
\caption{q-shock regular} \label{q-shockregular1-10-50}
\end{figure}

 \section*{Acknowledgments} One of the authors (SN) was partially supported by National
 Scholarship of the Scientific and Technological Research Council of Turkey (TUBITAK). This work was supported partially by
          Izmir Institute of Technology, Turkey.

\end{document}